# Wind Power Transmission System Integration — a Case Study of China Wind Power Base

Jianxue Wang[a], Shutang You[b], Xingzhong Bai[c], Mingqiao Peng[c]
a.  School of Electrical Engineering, Xi'an Jiaotong University, Xi'an, Shaanxi, 710049, China
b.  University of Tennessee, Knoxville. Knoxville, TN 37909, USA
c.  Northwest Branch of State Grid Corporation of China, Xi'an, Shaanxi, 710048, China

**Abstract:**

Due to a series of supporting policies in recent years, China wind power has developed rapidly through a large-scale and centralized mode. This paper analyzes the two major concerns faced by China's wind power development: wind generation reliability and wind energy balancing. More specifically, wind farm tripping-off-grid incidents and wind power curtailment issues, which caused huge economical loss, are investigated in details. Based on operation experience of large wind power bases, technical recommendations and economic incentives are proposed to improve wind power integration and power grid reliability. As a summary and outlook of wind power development in China, this paper provides a reference on future wind power development for other countries.

1.  Introduction

How to drive the economy with clean and sustainable energy while maintaining grid reliability is a huge challenge faced by the world's power grids. On the Climate Change Conference in Copenhagen 2009, the State Council of China pledged that in 2020 China will reduce greenhouse gas emissions per unit of GDP to 55-60% of that in 2005 [1, 2]. Among all types of renewable energy in China, wind power has the greatest potential for large-scale development. According to a plan issued by China, wind will be one of five major power sources and provide 17% of the national electricity demand by 2050 [3].

In some power grids, renewable energy is integrated into low-voltage distribution networks or strong grids with flexible synchronous generators to provide support, so three is no insurmountable obstacles in renewable energy development except for some system-level stability issues, for example, frequency stability. However, these favorable conditions and well-developed market environments do not exist for China's wind power. In fact, wind resource is mainly distributed in north and west China, far from electricity consumption centers. Therefore, wind power in China is being developed in a large-scale mode and integrated into weak transmission systems [4, 5]. Due to the characteristics of wind resources distribution and power grid conditions, wind power integration encounters serious difficulties, leading to a huge waste of resources [6]. In 2012, the curtailed wind power in China reached 20TWh or 20-30% of total wind power production. The value of curtailed wind power was about 10 billion RMB.

Although wind power is clean energy, its security issue is more concerned than conventional generators [7]. In most countries, wind power incidents occurred occasionally but few serious accidents happened. In U.K., for instance, wind power incidents almost happen once a day, but most incidents are faults of few Wind Turbine Generators (WTGs), which have little impact on systems [8, 9]. However, due to various factors from wind farms and power grids, the security issue of China's wind-integrated systems is very prominent. The large-scale wind power off-grid accidents in 2011 caused huge economic losses and made the wind power development almost stagnant.

China has attempted to take a series of policies and measures to ensure wind power safety and alleviate the difficulty of wind power integration. In this paper, the challenges confronted by China's wind power are elaborated. This paper also analyzes the current polices of wind power integration in China, and provides some detailed recommendations. The remainder of this paper is organized as follows. Section 2 provides the overall situation of wind power development in China. Section 3 gives a detailed description of the development situation and problems confronted by large-scale wind power bases in China. A statistical analysis of wind power accidents in Jiuquan Base and other bases is presented Section 4. Section 5 describes China's policies on wind power safety. Section 6 provides policy recommendations on China's wind power security and integration issues. Conclusions are provided in the last section.

## 2. The overall situation wind power development in China

### 2.1 Data of China's wind power development

Fig. 1 shows the wind capacity growth in China during 2001 to 2012. From 2000 to 2003, the annual growth rate of wind power installed capacity was about 30%. Since 2005, there has been a very rapid growth owing to the incentive role of the Renewable Energy Law [10]. From 2005 to 2010, the installed wind capacity in China nearly doubled every year. It can be seen that since 2011, China's wind power has entered a period of slow growth, the reason of which will be described in detail later in this paper.

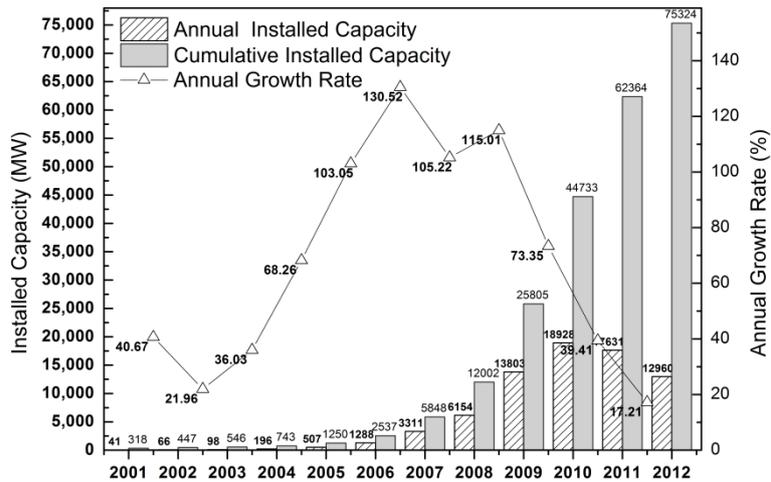

**Fig. 1.** Wind capacity growth in China 2001-2012. (Source: CWEA [11])

The majority of wind power in China is developed in a large-scale and centralized mode. In the document "Chinese Wind Power Development 12th Five-Year (2011-2015) Plan", China plans to build eight 10GW-scale wind power bases, which are located in seven provinces in North and East China. As shown in Fig. 2, wind resources in these areas are rich. The average wind power densities in these locations are greater than 150W/m$^2$. In some locations, the wind power density even reaches 400W/m$^2$. In 2020, the eight bases will account for 75% of the total installed wind power capacity in China.

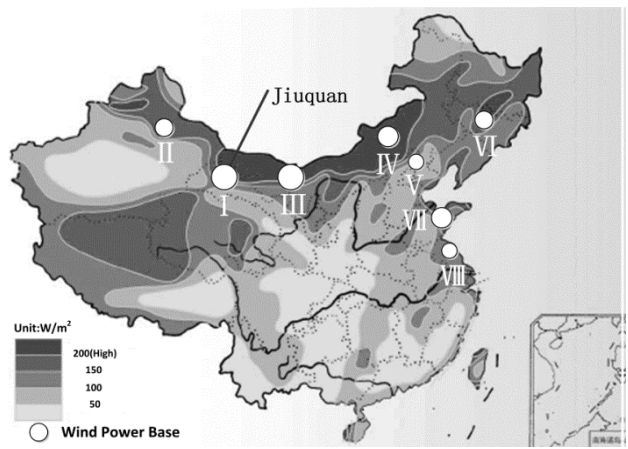

**Fig. 2.** Distribution of wind resource and the eight wind power bases. (Source: Liu [12])

Each wind power base is composed of dozens of wind farms which are geographically adjacent to each other. Since the installed capacity of most wind farms is more than 200MW, the aggregate capacity of a wind power base is usually extremely large, being more than 10GW. Table 1 shows the installed and the expected capacity of eight wind power bases during 2008 to 2020.

Table 1. Capacity of wind power bases in China during 2008 to 2020. (Source: CWEA [11])

| Provinces (Base No.) | Wind Power Capacity (GW) | | | | | |
|---|---|---|---|---|---|---|
| | 2008 | 2009 | 2010 | 2011 | 2012 | Expected 2020 |
| Gansu (I) | 0.60 | 2.00 | 5.16 | 5.16 | 6.48 | 22.00 |
| Xinjiang (II) | 0.58 | 1.00 | 1.36 | 2.32 | 3.31 | 11.00 |
| Inner Mon. (III,IV) | 3.65 | 9.20 | 13.86 | 17.59 | 18.62 | 59.00 |
| Hebei (V) | 1.11 | 2.79 | 4.79 | 6.97 | 7.98 | 14.00 |
| Jilin (VI) | 1.07 | 2.06 | 2.94 | 3.56 | 4.00 | 21.00 |
| Shandong (VII) | 0.56 | 1.22 | 2.64 | 4.56 | 5.69 | 11.00 |
| Jiangsu (VIII) | 0.65 | 1.10 | 1.60 | 1.98 | 2.37 | 10.00 |

## 2.2 Problems faced by China's wind power development

In China, most of the wind power bases are far away from load centers. For instance, the distances between Base II in Xinjiang and load centers located in Central and East China are 2,500km and 4,000km, respectively. Therefore, a strong and high-voltage electric network is needed to transmit a large amount of wind power. In fact, for GW-class wind power, the existing networks' capability is insufficient. However, electric network expansion requires huge investment and a long construction time.

More specifically, wind power bases have the following technical characteristics: large capacity, high voltage level, long transmission distance. Moreover, existing transmission networks are difficult to meet the demand of large-scale wind power development.

In terms of technical standards, China did not keep pace with the wind power development in the past several years. In order to mitigate wind power's impacts and guide development, Chinese regulatory entities have issued several technical standards. However, due to lack of an explicit and unified technical standard, these various technical standards brought about confusion. Rapid but unordered wind power development has huge impacts on the power system operation in China [13].

Unsuitable management policies also hinder the development of wind power in China. Different from technical standards, management polices need to consider the characteristics of local power system conditions. Therefore, specific management policies are usually formulated by the local gird operators. Management polices usually contain the assessment of following fields: wind power prediction accuracy, technical assessment on grid-integration and routine management. Since developing a suitable management policy needs to consider many factors, it is usually not an easy task.

In order to conduct a detailed study of China's wind power bases, this paper take Jiuquan wind power base (hereinafter referred to as Jiuquan Base), denoted by Base I in Fig. 2, as a representative.

## 3. Analysis on the challenges confronted by Jiuquan Wind Power Base

*3.1 Basic situations of Jiuquan Wind Power Base*

Jiuquan Base, as a lead wind power base, is located in the northwest part of Gansu Province, in a valley called Hexi Corridor. One of the salient features of Jiuquan Base is the centralized mode of wind power generation. By the end of 2011, Jiuquan Base consisted of 32 wind farms and its capacity is 5166MW. Moreover, its long term prospect capacity is more than 30GW [14].

The grid in Hexi Corridor also has a long and narrow structure. Wind farms are integrated into the central-west part of the electric network. The grid integration of Jiuquan Base has the following typical features. a) Large-scale, centralized mode. b) Far away from load centers. The distance between Jiuquan Base and Lanzhou City, the nearest load center, is 800km. c) High voltage level: the wind farms are integrated into the grid through 110kV substations, and further through 330kV and by 750kV substations. The large amount of wind power is transmitted through 750kV lines to load centers. The grid structure in Hexi Corridor is shown in Fig. 3.

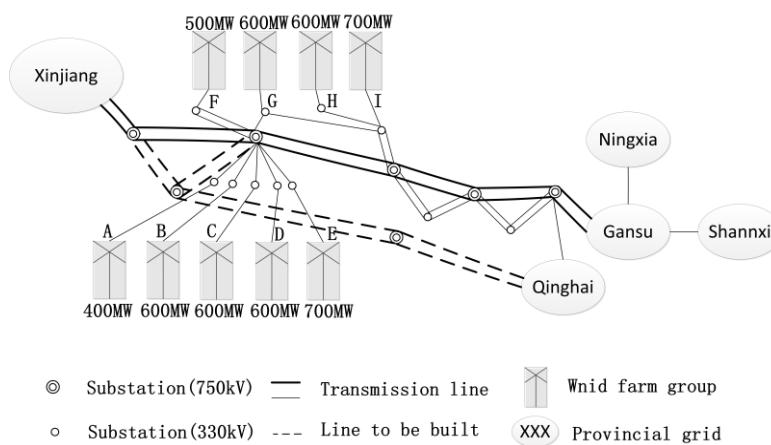

**Fig. 3.** The grid structure in Hexi Corridor.

*3.2 Unfavorable factors of Jiuquan Wind Power Base*

The key characteristic that makes wind power different from other conventional power sources is the volatility of available wind resource. Because of this volatility, the resulting problems include frequency and voltage stability, insufficient transmission capacity, power quality, grid security, and generation scheduling, etc. [15-20]. In addition to these inherent defects, some unfavorable factors faced by Jiuquan Base in China are further described as follows.

### 3.2.1 High penetration and simultaneity rates

In Gansu Province, most of the wind generation capacity (about 94.7%) is installed in Jiuquan Base. Since the system load is not large, the wind power penetration rate is very high. Wind power accounts for one fifth of the total generation capacity in Gansu. In some periods, wind power generation was larger than hydro power generation, becoming the second largest power source except for thermal power. For instance, on April 22$^{nd}$ 2011, wind power generation accounts for 21% of the daily generation in Gansu Province [21].

The simultaneity rate is used to describe the correlation between different wind farms' output. The simultaneity rate of a cluster of wind farms can be briefly defined as: during a certain period of time, the ratio between the maximum power output of the wind farm cluster and the sum of each wind farm's maximum output in this cluster. Table 2 shows the simultaneity rates of three wind farm clusters in Jiuquan Base. It can be noted that the simultaneity rate of Cluster 1 is smaller than those of Cluster 2 and 3. Combined with Fig. 3, the reason can be inferred as that wind farms in Cluster 1 are relatively more geographically scattered. However, on the whole, all the simultaneity rates in Table 2 are very high, reflecting a significant correlation between the outputs of wind farms in Jiuquan Base. Due to high simultaneity rates, the volatility of wind power output is aggravated.

Table 2. The simultaneity rates of wind farm clusters in Jiuquan Base. (Source: Wang [14])

| Cluster (group included) | Capacity (GW) | Voltage level of substation (kV) | Simultaneity rate |
|---|---|---|---|
| 1 (A- E, G) | 3.3 | 750 | 0.84 |
| 2 (F) | 1.3 | 330 | 0.91 |
| 3 (H, I) | 0.5 | 330 | 0.92 |
| All (A-I) | 5.1 | | 0.81 |

### 3.2.2 Adverse correlation between wind power output and load curves

Generally speaking, wind power output is unevenly distributed on a daily time scale. The output in Jiuquan Base is usually the largest during 0-6h, accounting for 44% of the daily total output. However, during this period, system load happens to be the lowest. This inverse correlation between the output curve and the load curve increases the difference between the peak and valley of the load curve, as shown in Fig. 6. This indicates that when wind power is added, the system needs more reserve capacity to maintain a supply-demand balance.

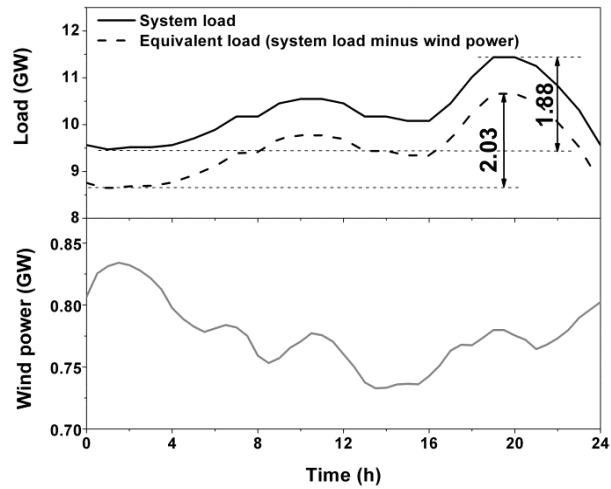

**Fig. 4.** The average daily wind output curve and the typical daily load curve in Gansu (2011). (Source: Zhang, Zhou [22])

### 3.2.3 Lag of grid construction

Due to lack of a unified and coordinate development plan in Jiuquan Base, transmission construction lags behind wind farm construction. Since the construction of wind farms is generally much faster than that of electrical networks, lack of wind transmission capability will exist in Jiuquan Base in the foreseeable future. The transmission bottleneck is also a common problem in China's large-scale wind power development.

Among all of the problems brought by these unfavorable factors, off-grid accidents and the wind power integration difficulty are the most critical.

### 3.3 Large-scale off-grid accidents

The three characteristics mentioned above not only increase system risk, but also make the consequences more serious once a large-scale wind farm off-grid accidents occurs, due to following reasons.

First, the close electrical link between wind farms extends the scale of accidents. As mentioned above, all the wind farms in Jiuquan Base are closely linked to each other electrically due to the centralized grid-integration mode. Therefore, the voltage sag caused by a short circuit of one WTG or one wind farm will greatly affect the voltage level of other WTGs and other wind farms, leading other WTGs trip off-grid. In the development processes of large-scale wind power accidents, such a chain effect is very common.

Second, the high simultaneity rate reinforces interactions and mutual influences between wind farms, making large-scale accidents more likely to occur. Since the grid operation mode is greatly depended on the output sum of the wind base, if the variation of each wind farm's output is highly simultaneous, the system's security margin would be rather small. Moreover, the robustness of the entire system is also relatively poor. The high penetration rate also aggravates the adverse effects of accidents because the system contains a large proportion of relatively "unreliable" WTGs. For instance, in a system with high wind power generation, a system fault will result in many WTGs' tripping off, leading to more generation capacity loss[23].

Third, the high volatility of wind power and the lag of grid construction make reserve capacity scheduling a difficult task for system operators. When wind speed decreases rapidly, the system faces great risks due to insufficient fast-response spinning reserve. Under this circumstance, if a wind power accident occurs, system risk will further increase. In addition, the lag of grid construction leads to that the lost wind power generation cannot be replaced by generation in other areas, so the accidents of wind power result in more load curtailment and more severe frequency fluctuations, which in turn cause more WTGs to trip off grid.

*3.4 Wind power integration difficulty*

The wind power integration difficulty is another important problem faced by Jiuquan Base besides wind power accidents. Annual Utilization Hour (AUH) is used here to reflect the utilization level of wind energy. The number of AUH reflects the utilization efficiency of WTGs. This indicator directly affects economic benefits of wind farms. In 2012, the number of wind power AUH in Jiuquan Base is 1,645h, compared with 4,670h and 3,540h respectively for thermal and hydro power in Gansu Province. The number is also slightly below the average number of wind power AUH in the northwest regional grid (1,853h) and nationwide (1,890h). The small number of AUH means that much wind energy is curtailed.

In many developed countries in Europe and America, flexible generators (e.g. gas-fired and pumped-storage hydropower) account for a relatively high proportion in the generation mix. These generators can well balance wind power fluctuations. For example, flexible generators in Spain account for 45.4% of its total generation capacity in 2012, providing sufficient regulation capacity to accommodate wind power variation. Some countries with high wind power are supported by powerful trans-national and trans-regional interconnected power systems. Thus, the areas and resources to balance wind power are expanded. Denmark is a typical country that benefits from large-scale interconnected power systems. In Denmark, wind power generated 28.2% and 30% of annual electricity consumption in 2011 and 2012 respectively. Denmark power grid is not only a part of Nordic power system, it also connects to European mainland system. The strong interconnected system provides sufficient balancing capacity to guarantee a high proportion of wind power in Denmark. In addition, as a member of the Nordic power market, Denmark is in a well-developed power market environment containing real-time electricity price, ancillary service trading and demand side management [24], which alleviate the operation difficulties owing to a high wind power proportion.

The wind power integration problem in China is caused by a variety of factors. In terms of technology, the large amount of wind power curtailment is mainly due to lack of the transmission capability, insufficient ramping down ability of large coal-fired power plants during low-load hours, and the effects of wind power accidents. From the perspective of policy, the integration problem is caused by the following reasons. 1) Current policies lack reasonable incentive mechanisms to encourage the construction of peak-shaving power plants. 2) The approvals of power network interconnection projects progress very slowly. 3) The electricity price of the user side is fixed and unitary, so the demand side management is weak. 4) Thermal power companies lack motivation to reduce their output to allow wind farms to generate more wind power.

Since abandoning wind energy is uneconomic for both wind farms and society, measures should be taken to facilitate full utilization of wind power. Based on international polices and experience, as well as China's situations, some policy recommendations in above aspects are proposed in Section 6.3.

## 4. Synopsis of wind power accidents in China

*4.1 Statistics of China's wind power accidents*

Despite the booming construction of wind farms, the large number of accidents reveals many problems in development. In 2010, 80 accidents of wind farm occurred in China. Among them, accidents in which 100-500MW and more than 500MW wind power output were lost account for 17.5% and 1.3% of the total number of accidents respectively. However, in 2011, the two ratios increased to 28.0% and 6.2% respectively, and the total number of accidents more than doubled [25]. Fig. 5 shows the statistics of accidents in several wind power bases during 2010 and 2011. The data indicate that wind power security has become an increasingly important issue.

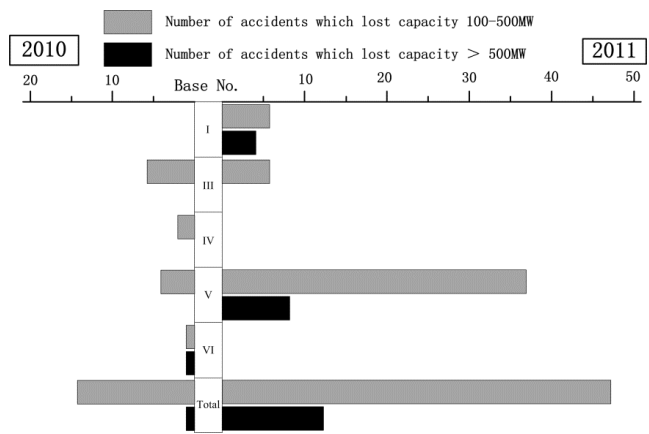

**Fig. 5.** Statistics of accidents in several wind power bases (2010-2011). (Source: SERC [26])

*4.2 Experience and analysis on wind power accidents*

During February to April in 2011, four large-scale off-grid accidents occurred in Jiuquan

Base. The lost wind power ranges between 568MW to 1535MW [25]. Table 4 shows the statistics of the four wind power accidents.

Table 3. Statistics of the four wind power accidents in Jiuquan Base in 2011. (Source: SGCC [27])

| Date | Initial cause | Capacity of off-grid WTGs (MW) | | | Off-grid proportion |
|---|---|---|---|---|---|
| | | Low voltage | High voltage | Frequency module defects | |
| Feb. 24th | Cable | 377.1 | 424.2 | 36.0 | 54% |
| Apr. 3rd | terminal | 445.5 | 103.0 | 19.5 | 29% |
| Apr. 17th | defects | 794.2 | 63.0 | 0.0 | 53% |
| Apr. 25th | Weather | 1,056.2 | 110.0 | 0.0 | 78% |

From accidents' occurrence to recovery, typical accidents' developing processes can be summarized to four stages as follows:

The first stage — the short-circuit fault stage (accident-causing stage). Ordinary short-circuit faults in wind farms result in low-voltage of WTGs' terminals. Typical short-circuit faults may be caused by installation quality defects of cable terminals, or common network faults. During short-circuit faults, the voltage value at PoC declines to about 75% of its initial value.

The second stage — the low-voltage off-grid stage (accident-formation stage). After the occurring of short-circuit faults, the speed of blades rises and WTGs absorb a lot of reactive power. If a WTG lacks in LVRT capability, it has to trip off-grid to protect its components from damage.

The third stage — the high-voltage off-grid stage (accident-development stage). After the clearance of fault and lots of WTGs' tripping off, if reactive power compensation devices are not regulated, reactive power becomes excessive gradually. Then the voltage at PoC increases, whose instantaneous value may reach 1.1 times of its rated value. Thus the high voltage protection of WTGs is activated and WTGs trip off-grid, thus in turn causing reactive power more excessive and more off-grid WTGs due to high voltage.

The fourth stage — the recovery stage. Guided by system operators, wind farm operators adjust reactive power compensation devices to adjust voltage to its normal level. Then WTGs gradually resume operation.

The initial causes of accidents are usually ordinary short-circuit faults in wind farms or external grids. Faults result in a low voltage drop, which may not be serious enough to lead to conventional generators' tripping off-grid. However, this voltage drop can cause WTGs without LVRT capability to trip off. Accident statistics also verified that low voltage causes most WTGs' tripping off, indicating that lacking of LVRT capability directly makes the fault of a single wind farm to evolve into large-scale wind power accidents. High-voltage off-grid, which occurs in the accident scale expansion stage, is due to imperfect adjustment of reactive power compensation equipment.

Having experienced these accidents, system operators summarized some anti-accident experience, which can be divided into the following aspects:

1) To prevent accident occurrence: quality management of devices' installation; maintenance management.
2) To minimize impacts: mandatory integration technical requirements (e.g. LVRT capability); relay protection devices management; reactive power compensation devices management.
3) Quick recovery: fast report of accidents' situations to grid operators; allocating system resources to reduce impacts of accidents.

In all the above measures, improving the LVRT capability is the important critical one. Operators' experience also proves this point. For instance, six WTGs in Group E (in Fig. 3) were retrofitted in the LVRT capability ahead of time, and during the accident on April 17$^{th}$, these retrofitted WTGs operated continuously, while other WTGs tripped off. From the above accidents' development process, it can be noted that lack of the LVRT capability also expands accidents' scale and impact. Therefore, in order to reduce system risks, LVRT capability management should be included in as a basic requirement. In addition, reactive power compensation devices, frequency module, and other hidden danger should be regularly checked to minimize accident impacts.

In short, wind power accidents are caused by inferior technologies of wind farms from a technical perspective and imperfect policies from a management perspective. Therefore, the solution should also cover the two aspects. The first is the refinement and implementation of technical standards. This will not only ensure equipment to be grid-friendly, but also provide the technical operation basis for wind generators. The second is providing wind farms economic incentives to improve safety management. Section 5, 6.1 and 6.2 will discuss these two aspects further.

## 5. Policies and measures to avoid large-scale off-grid accidents

In order to ensure system security, reliability and power quality, almost all developed countries with considerable wind power have special grid codes for wind power and they are strictly enforced. Denmark is the first country to formulate grid code for wind power, which went through many modifications and is highly detailed in its current version. Denmark's wind power grid codes are different for wind power being integrated into distribution and transmission networks. In Spain, the grid code has detailed requirements on WTG's control performance. Each wind farm should submit its technical test certification before connecting to girds to confirm its compliance with grid codes. After integration, system operators assess wind farms' operation performance and apply a bonus/penalty mechanism according to assessment results.

Combining experience of other countries and the technical investigation result of the four severe accidents of Jiuquan Base, the Chinese electricity regulatory agency — SERC took the two following two measures to prevent wind power accidents [25].

On the technical side, China formulates the new grid code —"Wind Farm Connecting Power Systems Technical Regulations", to replace the previous code "Technical Rule for Connecting Wind Power Plants with the Power Grid" issued in 2005 and the SGCC code "Wind Farm Connecting Grid Technical Requirements" [28]. Due to China's large-scale centralized wind power, the new national standard focuses on large-scale wind bases whose grid-integration voltage level is 110kV or more. In addition, compared with enterprise standards, the new standard is more authoritative, impartial and applicable.

On the management side, wind farms take four measures: LVRT capability retrofit, reactive power compensation device management, and protection and automation system management, diagnosis of cable terminals' defects. Meanwhile, system operators provide assistance in the implementation process of above measures. Regulation entities also regulate grid companies and ensure the fairness in dispatching wind power.

The following section provides technical details of the above policies.

*5.1 Comparison study on LVRT capability requirements*

In most grid codes, the LVRT capability is an important issue in wind farm integration. The LVRT capability ensures that when the voltage suddenly sags, a WTG operates continuously for a period of time before tripping off. Moreover, in high wind power penetration scenarios, the LVRT capability can avoid a large number of WTGs tripping off simultaneously, which significantly improves system reliability [29].

*5.1.1 Basic LVRT requirements*

The LVRT curve, representing the basic requirements on LVRT capability, is comprised of the instantaneous voltage sag line, the minimum voltage duration line, and the voltage restoring line. The national standard issued in 2005 did not give any requirements on LVRT capability. Due to the rapid development of wind power, the SGCC enterprise standard Q/GDW 392-2009 makes specified requirements on LVRT capability for the first time. Fig. 8 shows a comparison between LVRT capability requirements in China and some developed countries. As shown in Fig. 8, the voltage recovery process is faster in Chinese national standard than SGCC enterprise standard; therefore, the WTG's low voltage withstand capability is lower in the Chinese national standard. The national standard is a compromise which simultaneously considers the interest of wind farms and grid companies. In addition, from a worldwide perspective, the Chinese national standard is a moderate one while the SGCC enterprise standard is relatively strict.

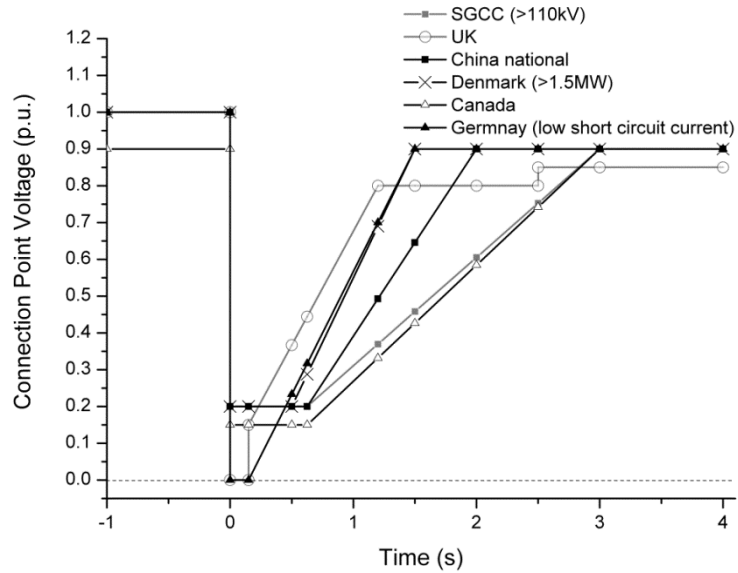

**Fig. 6.** Comparison on LVRT requirements between China and some developed countries. (Source: AESO [30]; Energinet [31]; E.ON Netz [32]; NEA [33]; NGET [34]; SGCC [28].)

*5.2 Implementation of polices and corrective measures*

Implementation of prevention measures requires a lot of resources. To illustrate its necessity, the cost of prevention measures and economic cost of accidents should be compared. Cost of prevention measures are mainly the cost of LVRT capability retrofit which is about 100,000 RMB/MW [35]. The number of retrofit-needed WTGs is about 2,500 with an aggregate capacity of 3,700MW (of the remaining WTGs, 14.0% are too old to be retrofitted, and 18.3% already possess LVRT capability). Thus，the total cost of prevention measures is about 370 million RMB.

The economic cost of accidents can be estimated through the commonly used method in evaluating lost load cost. This method estimates accident cost through multiplying the following values: duration of wind power loss (h), wind power loss (MW), cost of lost load (RMB/MWh), and the expected number of accidents during the whole life of wind farms. The life of wind farms is assumed to be 20 years, and four accidents are assumed to occur each year. Each accident's loss is the average of the four accidents in Section 4.1. The average duration of accidents is 10 minutes, and the cost of per unit load loss is 30 RMB/kWh. Finally, the obtained economic cost of accidents is 1,987 million RMB.

This shows that the cost of prevention measures is much less than the economic cost of accidents. Thus, the prevention measures are reasonable economically. LVRT retrofit has several optional solutions, whose advantages and disadvantages are shown in Table 6. These optional techniques can be used in combination to achieve better LVRT performance.

Table 6. LVRT remediation techniques comparison.

| Remediation technique | Advantages | Disadvantages |
|---|---|---|
| Crow-bar resistance | Simple and low cost | Severe transient process may occur |
| Control strategy improvement | Low cost; improve transient response | Cannot ride through severe voltage drops |
| Reactive power compensation device | Already in practical application and providing voltage support | Relatively high cost |
| Energy storage device | Good performance | High cost and no practical application yet |

## 6. Policies outlook

In order to maintain the sustainable development of wind power, further measures to promote wind power development are critically important. In the following sub-sections, several policy recommendations are provided. These recommendations mainly focus on two issues: one is ensuring the long-term safety operation of wind power in power systems; the other is promoting wind power integration.

*6.1 Perfecting and strict implementation of technical standards*

For security consideration, the newly installed WTGs in China must meet the national standard, which includes some other technical requirements in addition to LVRT capability and operation adaptability, as shown in Table 8. If one or more items of a wind farm does not comply with the standard, it cannot be permitted to connect to grids.

Table 8. Summary of key requirements in the 2011 national standard. (Source: NEA [33])

| Aspects | Requirements |
|---|---|
| LVRT capability | Stay on grid when voltage changes within the LVRT capability requirement; |
| Operation adaptability | Keep running when frequency and voltage vary within the allowable ranges; |
| Active power | Smooth adjustment ability; Output fluctuations within the allowable range; |
| Reactive power | Maintain and adjust voltage at PoC; Generate self-sufficient reactive power; |
| Wind forecasting | Providing 0-24h (short-term) and 15min-4h (ultra-short-term) wind power forecast. |

Due to different system structures and rapidly developing WTG manufacture technology, there are many integration technical standards that are currently used throughout the world. In developed countries, wind power integration standards are more comprehensive than Chinese standards and cover a variety of technical aspects. In terms of framework, standards in China and developed countries are generally the same. In these standards, requirements on wind power integration mainly focus on four aspects: LVRT capability, active power, reactive power, and operation adaptability [36, 37]. However, the level of detail in the Chinese standard is far less than those in developed countries. There are many special cases that are included other standards but not in the Chinese standard. Therefore, China should make the standard more detailed and consistent with the grid situation in China.

Besides grid-integration permission of new WTGs, there are some more measures that grid companies can take. First, it is recommended that system operators should supervise and ensure that reactive power compensation devices and other protective devices in wind farms are put into operation. Wide-area or local monitoring systems should be used to ensure the compliance of physical facilities' behaviors are consistent with models. Second, the system operators should cooperate with wind farms design, construction, and operation companies to formulate effective wind farm control plans according to specific situations.

*6.2 Recommended policies for wind power integration*

After solving the security problem, a key issue is utilizing more wind energy, i.e. wind power integration. Objectively speaking, wind power integration is not only a technical issue, but also a policy issue that should be solved by coordination of interests. In China's wind power sector, the fixed tariff cuts off the economic signal. In order to improve wind power integration in China, further incentive policies are needed. The policies should be formulated from a systematic and comprehensive perspective which covers power plants, grid companies, and customers. In national power system, the optimal allocation of overall resources can help to solve the wind power integration problem. Therefore, some policy recommendations are as follows.

First, large-scale wind power needs to be transmitted inter-provincially by grid companies. Because of the high penetration rate and the limited load level of a single provincial grid, thermal power generation will be severely limited and become very uneconomical. Moreover, wind power volatility needs more reserve capacity to balance, the amount of which is too large to be provided by a single provincial grid. Therefore, a strong enough inter-provincial power network should be constructed and an inter-provincial transmission and trading mechanism should be developed. As a basis of inter-regional wind power trading, high voltage transmission networks' construction should be strengthened, thus forming a wind power transporting channel between wind resource-rich regions and load centers.

Second, China should increase subsidies for grid companies to encourage them to further exploit potential integration capability of wind power. Since gird companies can take some effective measures to integrate more wind power, such as changing system operation modes, encouraging traditional generators to set aside more reserve capacity through economic measures, and selling excessive wind power to other grid companies, it is obvious that grid companies is an important role in the wind power integration system. From the perspective of grid companies, the electricity prices paid to wind farms and other conventional power plants are exactly the same. There is no economic difference between the wind generated electricity and thermal electricity for power grid companies, despite the fact that wind power is generally more unreliable. Therefore, grid companies are reluctant to integrate more wind power. The incentive method is to provide grid companies subsidies for the amount of wind power integration that exceeds beyond a basic level, thus linking wind power with the resource allocation mechanism of the whole system. This policy can compensate for the deficiencies of fixed wind power prices.

Third, the wind power consumption quota system should be formulated, in order to set wind power integration as a mandatory task. This quota system should clarify responsibilities and obligations of local regulatory entities, grid companies, and wind farms. For better implementation of this policy, a wind power trading mechanism should be developed and each region's wind power consumption proportion should be set. Developed regions should be economically motivated to accept more wind power. Regulatory entities can improve the energy price system to better reflect the scarcity and environmental costs of different energy sources. There are also some other measures to improve the competitiveness of wind power, such as energy tax adjustment.

Lastly, local peak-shaving capacity should be further developed, such as gas-fired power plants and demand side management. Because of the volatility and intermittency of wind power, generation from other power plants is needed to balance wind power. Then the mixed generation is a stable power supply and can be delivered to distant load centers. Therefore, a certain number of thermal power plants should be constructed near Jiuquan Base to deliver more wind power. In addition, economic incentives should be further added to improve the compensation mechanism of thermal plants' peak-shaving for wind power. As for demand side management, heat-supply companies and heavy-load consumers should be economically coordinated with wind farms. The demand management measures include time-varying prices and interruptible load. These measures will further balance the fluctuations in wind power and promote wind power integration in a wider range of area.

## 7. Conclusions

Different from the distributed development mode, wind power in China develops in a large-scale and centralized mode. Without direct economic measures from electricity markets of developed countries, managing such large-scale wind power is a difficult task. Based on lessons from a series of accidents, the current policies in China mainly cover two aspects: standardization of technical requirements and implementation of retrofit projects. Results indicate that the existing policies are effective in preventing large-scale wind power accidents. This paper provides some policy recommendations to further improve the security and utilization of wind power. This information also serves as a reference for wind power development in other countries and regions.

## Acknowledgments

This work was supported by the National Natural Science Foundation of China (No. 51277141) and the National High Technology Research and Development Program of China (863 Program, No. 2011AA05A103).